%% file: FEIR.tex
\documentclass[sigconf,svgnames]{acmart}

\usepackage{booktabs} 
\usepackage{threeparttable}
\usepackage{colortbl} 
\usepackage{color}
\usepackage{textcomp}
\usepackage{soul} 
\usepackage{amsmath}
\usepackage{comment}
\usepackage{multirow}
\usepackage{algorithm} 
\usepackage{algpseudocode} 

\setcopyright{rightsretained}

\begin{document}
\title[Formula Embedding for MIR]{Preliminary Exploration of Formula Embedding for Mathematical Information Retrieval}
\subtitle{Can Mathematical Formulae Be Embedded like A Natural Language?}

\author{Liangcai Gao}
\affiliation{%
  \institution{Institute of Computer Science and Technology}
  \institution{Peking University}
  \city{Beijing} 
  \country{China} 
  \postcode{100871}
}
\email{glc@pku.edu.cn}

\author{Zhuoren Jiang}
\authornote{Zhuoren Jiang is the corresponding author}
\affiliation{%
  \institution{Institute of Computer Science and Technology}
  \institution{Peking University}
  \city{Beijing} 
  \country{China} 
  \postcode{100871}
}
\email{jiangzr@pku.edu.cn}

\author{Yue Yin}
\affiliation{%
  \institution{College of Information Science and Technology}
  \institution{Beijing Normal University}
  \city{Beijing} 
  \country{China}
  \postcode{100875}
}
\email{bnuyinyue@outlook.com}

\author{Ke Yuan}
\affiliation{%
  \institution{Institute of Computer Science and Technology}
  \institution{Peking University}
  \city{Beijing} 
  \country{China} 
  \postcode{100871}
}
\email{yuanke@pku.edu.cn}

\author{Zuoyu Yan}
\affiliation{%
  \institution{Yuanpei College}
  \institution{Peking University}
  \city{Beijing} 
  \country{China} 
  \postcode{100871}
}
\email{yanzuoyu3@pku.edu.cn}

\author{Zhi Tang}
\affiliation{%
  \institution{Institute of Computer Science and Technology}
  \institution{Peking University}
  \city{Beijing} 
  \country{China} 
  \postcode{100871}
}
\email{tangzhi@pku.edu.cn}




\begin{abstract}
 While neural network approaches are achieving breakthrough performance in the natural language related fields, there have been few similar attempts at mathematical language related tasks. In this study, we explore the potential of applying neural representation techniques to Mathematical Information Retrieval (MIR) tasks. In more detail, we first briefly analyze the characteristic differences between natural language and mathematical language. Then we design a ``symbol2vec'' method to learn the vector representations of formula symbols (numbers, variables, operators, functions, etc.) Finally, we propose a ``formula2vec'' based MIR approach and evaluate its performance. Preliminary experiment results show that there is a promising potential for applying formula embedding models to mathematical language representation and MIR tasks.
\end{abstract}

\settopmatter{printacmref=false, printccs=false}

%
%




\maketitle

\input{content/intro}
\input{content/method}
\input{content/conclusion}
\section{Acknowledgments}
 This work is supported by the projects of National Natural Science Foundation of China (No. 61472014), the Beijing Nova Program (Z151100000315042) and the China Postdoctoral Science Foundation (No. 2016M590019), which is also a research achievement of Key Laboratory of Science, Technology and Standard in Press Industry (Key Laboratory of Intelligent Press Media Technology). We also thank the anonymous reviewers for their valuable comments.
\bibliographystyle{ACM-Reference-Format}
\bibliography{refs} 

\end{document}

%% file: content/intro.tex
\section{Introduction and Motivation}
Mathematical formulae are important means for dissemination and communication of scientific information \cite{zanibbi2016ntcir}. They are used for calculation and definition. Mathematical Information Retrieval (MIR), namely searching for a particular mathematical formula, concept or object, is an important area in Information Retrieval (IR). The most common form of mathematical information is formula, which is also the most challenging part of MIR \cite{lin2014mathematics}.

Recently, neural network approaches, e.g., word embedding \cite{mikolov2013efficient}, are significant for textual content representation, which can discover distributed representations of words and capture the semantic similarity between the words. Meanwhile, various neural methods have been applied in textual information retrieval, and some of them have achieved significant performance improvements \cite{guo2016deep}. However, there have been few similar attempts at mathematical language related tasks. Naturally, given the success of neural network models in natural language related domains, we begin to investigate these following research questions:
\begin{itemize} 
    \item \textit{\textbf{RQ1.} Can we introduce neural representation approaches (e.g., word embedding) into ``Mathematical Language''?}
    \item \textit{\textbf{RQ2.} Can we use neural representation techniques to improve the performance of Mathematical Information Retrieval?}
\end{itemize} 

 Generally speaking, formula, described by mathematical language, has many qualities in common with natural languages: it has symbols, and grammatical rules for combining the symbols. However, there are significant differences between formula and natural language. 1) The mathematical symbols of a formula are ambiguous. For instance, a variable ``$x$'' could represent a lot of meanings, sometimes it cannot be understood without a written or spoken explanation. 2) Formulae always have recursive structures while natural language is usually linear in structure. 3) Different from plain text, formulae are highly structured and usually presented in layout presentations, e.g., \LaTeX{} or \textit{MathML}. These characteristics may lead to the difficulties of applying the natural language approaches to the field of mathematical formulae.
 
In this paper, to investigate the proposed research questions, first, following the word2vec approach, we design a ``symbol2vec'' method to learn the vector representations of formula symbols (numbers, variables, operators, functions, etc.) Second, we propose a ``formula2vec'' based MIR approach and evaluate its performance. Preliminary experiment results show that the representation performance of formula symbol embedding is promising and there is a potential for applying formula embedding approach to MIR tasks.

%% file: content/method.tex
\section{Formula Embedding for MIR}
\subsection{Formula Symbol Embedding} 

The ``distributional hypothesis'' states that terms that are used (or occur) in similar context tend to be semantically similar. This hypothesis has motivated the work of word embedding and has been proven true in natural language domain \cite{mikolov2013efficient,mikolov2013distributed}. We verified this hypothesis in a mathematical language context. 

\begin{table*}[h]
\centering
\caption{Examples of the closest formula symbols given the "symbol2vec" model training results.}
\label{tab:case}
\begin{tabular}{c c c c c c c c c}
\toprule
\textbf{Symbol} & \multicolumn{8}{c}{\textbf{Most similar formula symbols}} \\ \midrule
$\backslash$sin & $\backslash$cos & $\backslash$cot & $\backslash$tan & $\backslash$csc & $\backslash$arcsin & $\backslash$arctan & $\backslash$sec & $\backslash$sinh \\
$\backslash$epsilon~($\epsilon$) & $\backslash$delta~($\delta$) & $\backslash$mu~($\mu$) & $\backslash$Sigma~($\Sigma$) & $\backslash$varepsilon~($\varepsilon$) & $\backslash$hbar~($\hbar$) & $\backslash$sigma~($\sigma$) & $\backslash$pi~($\pi$) & $\backslash$rho~($\rho$) \\
$\{$array$\}$ &  $\{$matrix$\}$ &  $\{$vmatrix$\}$ &  $\{$Vmatrix$\}$ &  $\{$bmatrix$\}\backslash$n &  $\{$Bmatrix$\}$ & $\{$vmatrix$\}\backslash$n &  $\{$pmatrix$\}\backslash$n &  $\{$Bmatrix$\}\backslash$n\\
5 & 6 & 7 & 8 & 9 & 4 & 3 & 6$\backslash$n & 9$\backslash$n\\
a & b & c & l & m & c$\backslash$n & o & s & a$\backslash$n\\
Orange & BrickRed & Violet &$\backslash$color & Red & RedViolet & blue & Blue & Brown \\
= & $\backslash$approx~($\approx$) & $\backslash$ge~($\ge$) & $\backslash$equiv~($\equiv$) & $\}$ & - & $\backslash$neq~($\neq$) & $\backslash$ne~($\ne$) & $\backslash$limits \\
+ & - & 2 & 3 & \^{} & 4 & $\}\backslash$n & 1 & 5 \\ \bottomrule
\end{tabular}
\end{table*}

\begin{figure*}[h]\centering
 	\includegraphics[width=1.6\columnwidth]{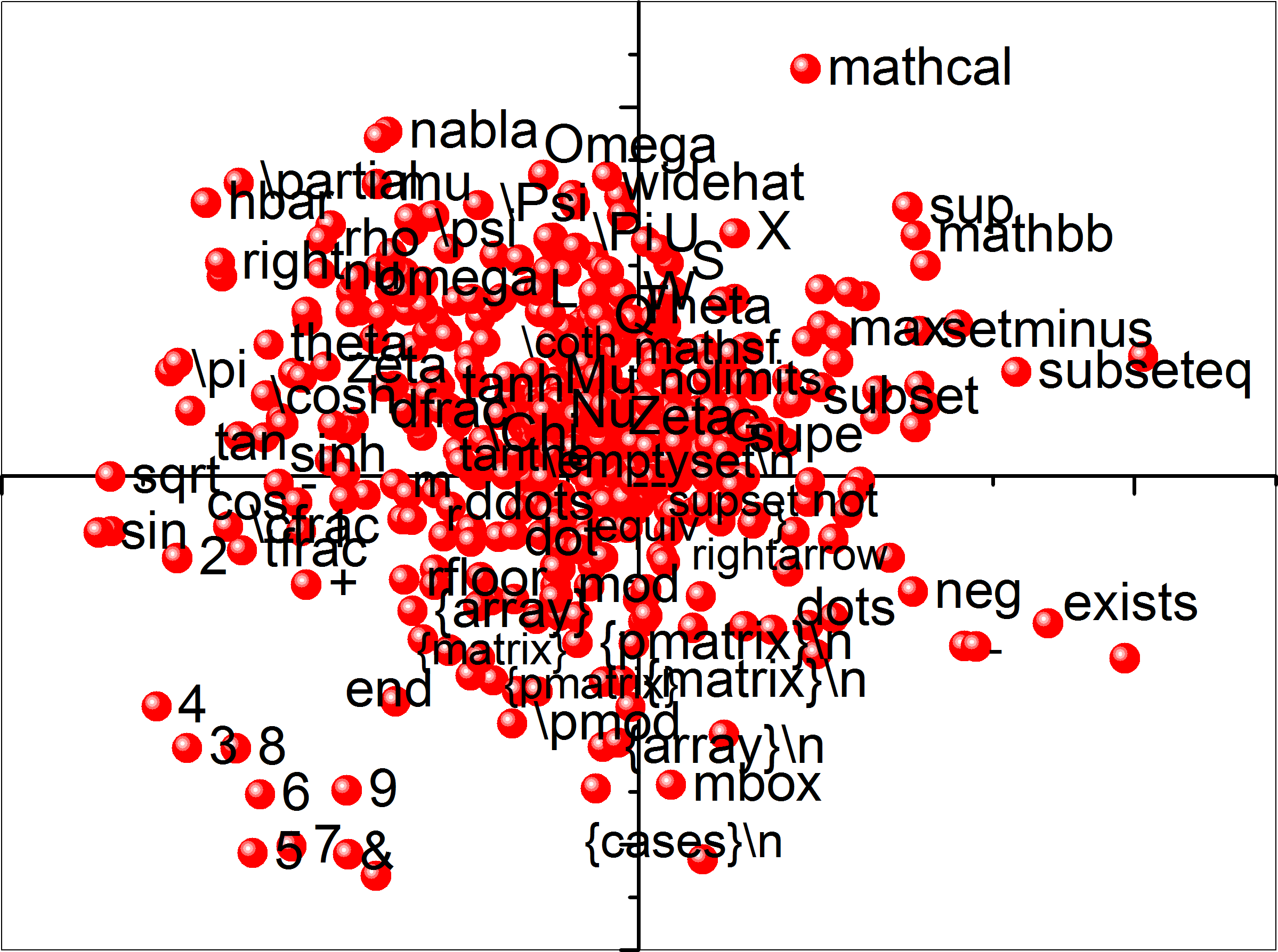} 
 	\caption{Two-dimensional PCA projection of the 100-dimensional vectors of formula symbols. (Note: to avoid overlap, the figure doesn't show all \LaTeX{} symbol labels.)}
 	\label{fig:graph}
\end{figure*}

In the preliminary experiment, we used a Wikipedia dump of July 30, 2014 to generate mathematical formula corpus. There were 358,116 raw formulae (in \LaTeX{} format). We then kept the formulae featuring at least two variables and three operators. Finally, there were 194,150 formulae left. A ``symbol2vec'' method was proposed following the CBOW architecture using negative sampling \cite{mikolov2013efficient}. One of the main challenges was how to convert formulae into minimal and meaningful \LaTeX{} terms (symbols). We utilized a formula tokenizer provided by \cite{deng2016you}. After tokenization, there were totally 892 \LaTeX{} formula symbols. The dimension of learned symbol embedding was 100. We calculated and ranked the most similar symbols for each symbol based on their cosine similarity, Table \ref{tab:case} listed 8 different examples. Furthermore, Figure \ref{fig:graph} visualized the learned embeddings.

Generally, the representation performance was promising. For instance, the representation vectors of numbers, variables or trigonometric functions were close to each other. Meanwhile, there were ``bad'' cases, for example, as Table \ref{tab:case} showed, the closest symbols to operator ``$+$'' were messy. The possible reason may be that formulae are highly structured, the ``semantic meaning'' of an operator symbol is not similar to its linear neighbor symbols (usually are numbers or variables).

The preliminary experiment results of "symbol2vec" revealed the several characteristics of mathematical formulae, which indicated the natural language embedding technologies (i.e., \cite{mikolov2013efficient}) could be potentially useful for formula embedding task. A non-linear formula tokenizer (i.e., tree-based) might be more helpful for finding the suitable context for a formula token.

\subsection{Formula Embedding Based MIR}
Typically, neural ranking models for textual information retrieval used shallow or deep neural networks to rank search results in response to a query. Following the general textual retrieval framework \cite{guo2016deep}, we proposed a ``formula2vec'' based MIR approach: 
$$ match(f_{1},f_{2}) = F(\Phi (f_{1}),\Phi (f_{2}))$$
where $f$ was a formula, $\Phi$ was a function to map each formula to a representation vector, and $F$ was the scoring function. 

\begin{figure}[h]\centering
 	\includegraphics[width=0.85\columnwidth]{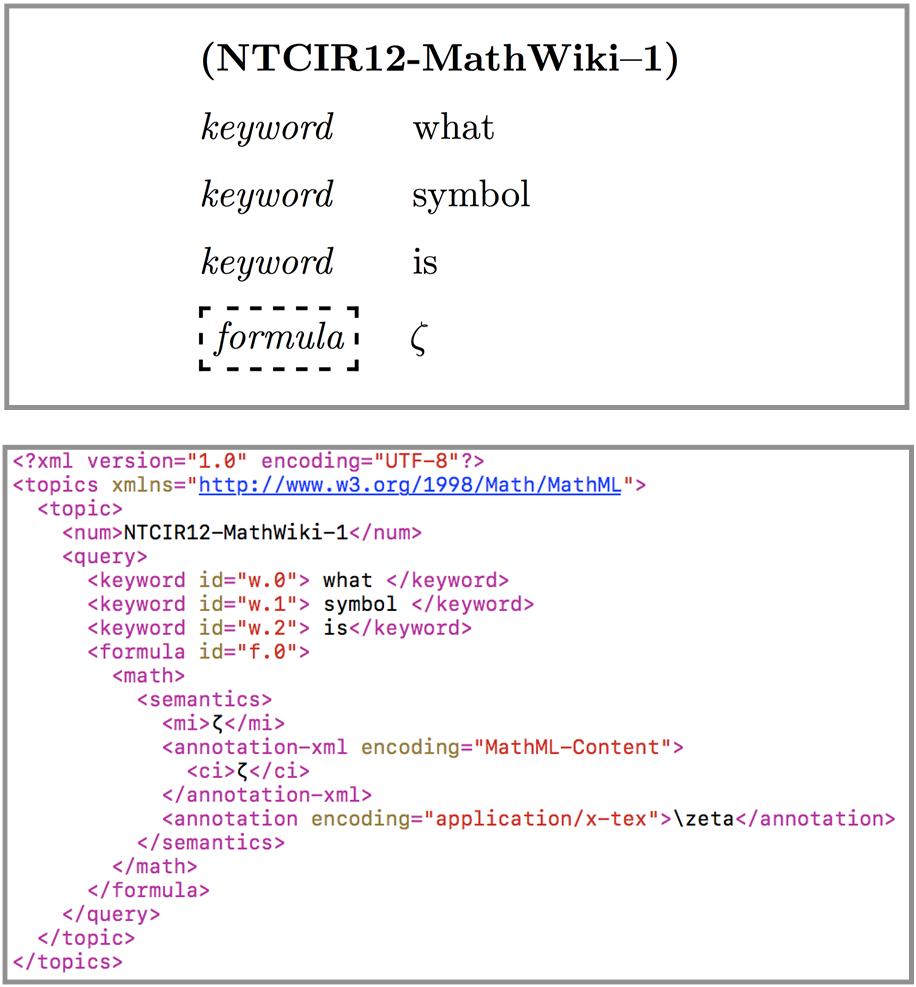} 
 	\caption{Query example of NTCIR-12 MathIR task and its corresponding XML description}
 	\label{fig:query}
\end{figure}

\begin{table*}[h]
\centering
\caption{Measures of mathematical information retrieval performance under different ranking methods (The dimension of ``formula2vec'' is 300, the $\alpha$ value for combined method is 4 )} 
\label{tab:expgroup}
\begin{tabular}{l  l  l  l  l  l  l}
\toprule
\textbf{Ranking} & \textbf{NDCG@30} & \textbf{NDCG@50} &\textbf{P@30} &  \textbf{P@50} & \textbf{MAP} & \textbf{MRR} \\ \midrule
Formula2vec & 0.0091 & 0.0089 & 0.0100 & 0.0067 & 0.0050 & 0.0572 \\ 
LM & 0.0739 & 0.0786  & 0.1111 & 0.0980 & 0.0646 &  0.3563 \\ 
\textit{\textbf{Formula2vec+LM}} & \textit{\textbf{0.0779}} & \textit{\textbf{0.0822}} & \textit{\textbf{0.1144}} & \textit{\textbf{0.1000}} & \textit{\textbf{0.0672}} & \textit{\textbf{0.3845}} \\ \bottomrule
\end{tabular}
\end{table*}

\begin{figure}[h]\centering
 	\includegraphics[width=1.0\columnwidth]{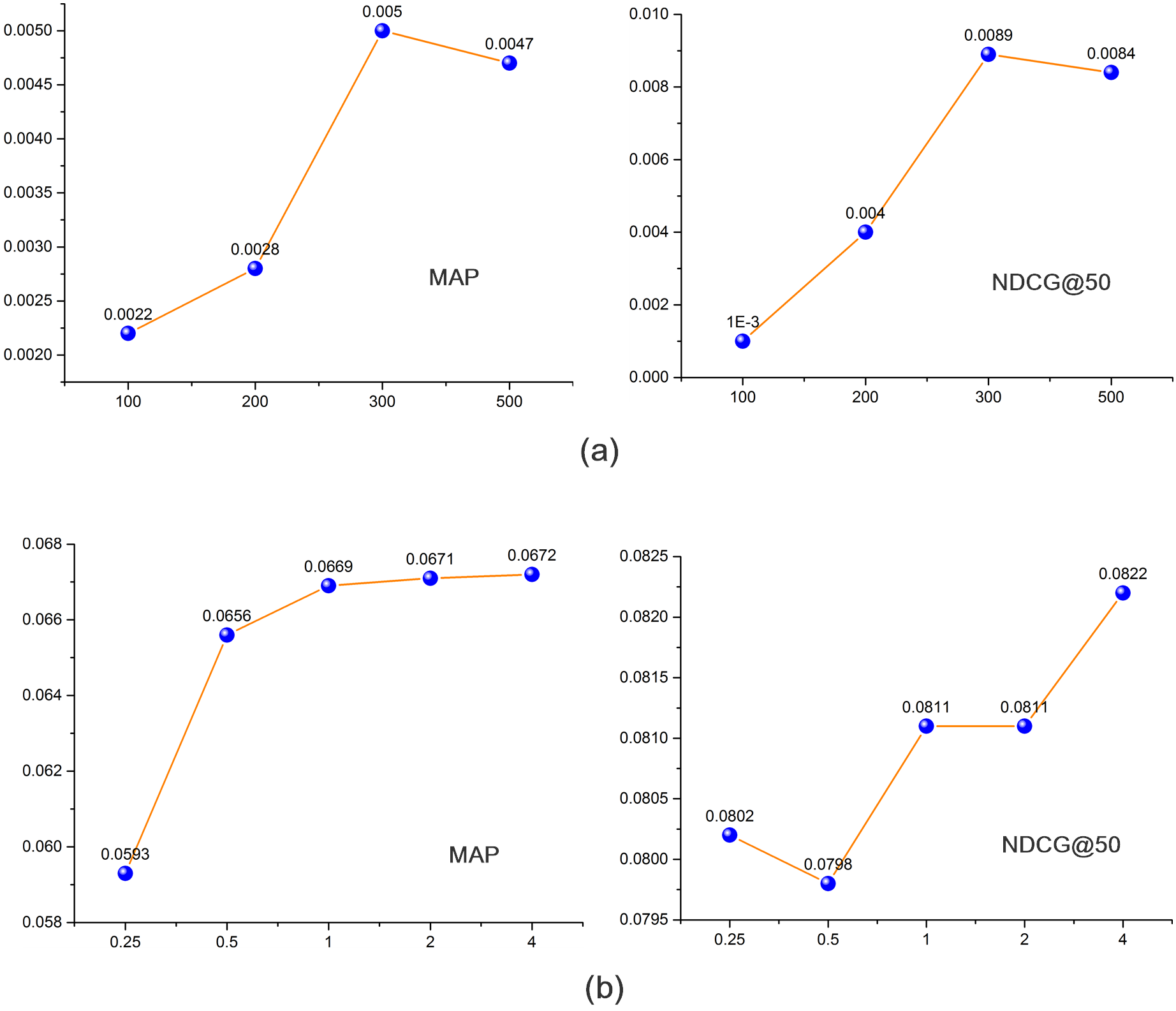} 
 	\caption{(a). Measures of ``formula2vec'' performance under different formula embedding dimensions; the y-axis represents the evaluation metrics, whereas the x-axis is the formula embedding dimension. (b). Measures of combined method performance under different $\alpha$; the y-axis represents the evaluation metrics, whereas the x-axis is the value of $\alpha$.}
 	\label{fig:tune}
\end{figure}

In the preliminary experiment, we used a Wikipedia evaluation collection released by NTCIR-12 MathIR task \cite{zanibbi2016ntcir}. NTCIR-12 MathIR task was designed for helping the undergraduate and graduate university students to locate the specific Wikipedia articles, browse math articles, or learn/review mathematical concepts and notations that they come across in the course of their studies. Figure \ref{fig:query} illustrated a typical mathematical query with its XML description. 
 
For ``formula2vec'' function $\Phi$, we utilized the Distributed Memory Model of Paragraph Vectors (PV-DM) \cite{le2014distributed}, which can learn distributed representations of formulae. For scoring function $F$, we simply employed the cosine similarity. As NTCIR-12 MathIR task only labeled the relevant score for Wikipedia page (not directly the relevance of formulae), we designed a ``formula2vec'' based Wikipedia page matching method. This method used the arithmetic mean value of all matching scores of formulae within a page, as the page relevant score. The pseudocode is described as Algorithm \ref{alo:learning}.

\begin{algorithm}[htbp]
\small
\caption{``Formula2vec'' Based Wikipedia Page Matching Method}
\label{alo:learning}
\begin{algorithmic}
\ForAll{Query $q \in$ QuerieSet $Q$}
    \State Initialize list [[Page $p$, PageScore $ps$]] psl
    \ForAll{page $p \in$ PageSet $P$}
        \State $ps$ = 0
        \ForAll{Formula $f_{q} \in q$}
            \State PageSimilarity $p\_simi$ = 0
                \ForAll{Formula $f_{p} \in p$}
                    $p\_simi +=$ CosineSimilarity $(f_{q},f_{p})$
                \EndFor
            \State $p\_simi /= n_{p}$ (Formula Number of A Page)
            \State $ps += p\_simi $
        \EndFor
        \State $ps /= n_{q}$ (Formula Number of A Query)
        \State $psl$.append([$p,ps$])
    \EndFor
    \State Rank($psl$)
\EndFor
\end{algorithmic}
\end{algorithm}
 
As Figure \ref{fig:query} shows, the queries of NTCIR-12 MathIR task contained both textual and formula information. However, the ``formula2vec'' based Wikipedia page matching method only utilized the formula information for ranking. To address this problem, in the preliminary experiment, we adopted a combined scoring method with a controlling parameter $\alpha$:
$$ C = \frac{F+\alpha T}{1+\alpha }$$,
in which, $F$ represented the ``formula2vec'' based Wikipedia page matching degree, $T$ represented the textual Wikipedia page matching probability using language model with Dirichlet smoothing \cite{zhai2001study}, $C$ was the final combined matching score, $\alpha$ controlled the contributions from textual information or formula embedding information. 

Table \ref{tab:expgroup} listed the performance under different ranking methods, in which, ``Formula2vec'' represented the formula2vec based  matching method, ``LM'' represented the textual matching probability using language model with Dirichlet smoothing, ``Formula2vec+LM'' was the combined method. The result indicated that: 1) the textual information is crucial for MIR performance. Comparing with the method only using formula embedding information, the language model method based on textual information can achieve a better performance. 2) the formula embedding information is helpful, by combining the ``formula2vec'' based matching degree and  textual matching probability, the combined method achieved the best performance. This observation proved the potential of the formula embedding approach applied in MIR tasks.

Figure \ref{fig:tune} (a) showed the ``formula2vec'' performance under different formula embedding dimensions. With the increase of dimension, the ranking performance was getting better, in terms of all evaluation metrics. When the dimension increased over 300, the performance could keep at a stable level (with a slight decline).

Figure \ref{fig:tune} (b) showed the MIR performance under different $\alpha$ values. From the results, we can conclude that, though both kinds of information were indispensable, comparing with the formula embedding matching information, the textual information should give more contributions in the combined method. This finding also proved the importance of textual information in MIR tasks.

%% file: content/conclusion.tex
\section{Analysis and Conclusion}
In this study, we explored the formula embedding approaches in mathematical information representation and MIR tasks. The preliminary exploration shows that: 
\begin{itemize} 
    \item The neural representation models can be applied for ``Mathematical Language'', the representation performance is promising.
    \item The characteristics of mathematical formula may lead to the poor representations of some specific symbols (i.e., operator ``$+$''). Instead of using the linear neighbor, we need a more advanced technology to detect the suitable ``context'' for a formula symbol (token).
    \item The formula embedding approach has potential for mathematical information retrieval tasks. By combining the textual information, the ``Formula2vec+Language model'' method can achieve the best retrieval performance.
    \item There is still a huge room for improvement of the formula embedding models.
\end{itemize} 
In the future, we will adopt a more suitable formula tokenizer (i.e., tree-based) for formula tokenization and investigate more sophisticated formula representation models to enhance the MIR performance.


%% file: FEIR.bbl

\begin{thebibliography}{00}


\ifx \showCODEN    \undefined \def \showCODEN     #1{\unskip}     \fi
\ifx \showDOI      \undefined \def \showDOI       #1{{\tt DOI:}\penalty0{#1}\ }
  \fi
\ifx \showISBNx    \undefined \def \showISBNx     #1{\unskip}     \fi
\ifx \showISBNxiii \undefined \def \showISBNxiii  #1{\unskip}     \fi
\ifx \showISSN     \undefined \def \showISSN      #1{\unskip}     \fi
\ifx \showLCCN     \undefined \def \showLCCN      #1{\unskip}     \fi
\ifx \shownote     \undefined \def \shownote      #1{#1}          \fi
\ifx \showarticletitle \undefined \def \showarticletitle #1{#1}   \fi
\ifx \showURL      \undefined \def \showURL       #1{#1}          \fi
\providecommand\bibfield[2]{#2}
\providecommand\bibinfo[2]{#2}
\providecommand\natexlab[1]{#1}
\providecommand\showeprint[2][]{arXiv:#2}

\bibitem[\protect\citeauthoryear{Deng, Kanervisto, and Rush}{Deng
  et~al\mbox{.}}{2016}]%
        {deng2016you}
\bibfield{author}{\bibinfo{person}{Yuntian Deng}, \bibinfo{person}{Anssi
  Kanervisto}, {and} \bibinfo{person}{Alexander~M Rush}.}
  \bibinfo{year}{2016}\natexlab{}.
\newblock \showarticletitle{What You Get Is What You See: A Visual Markup
  Decompiler}.
\newblock \bibinfo{journal}{{\em arXiv preprint arXiv:1609.04938\/}}
  (\bibinfo{year}{2016}).
\newblock


\bibitem[\protect\citeauthoryear{Guo, Fan, Ai, and Croft}{Guo
  et~al\mbox{.}}{2016}]%
        {guo2016deep}
\bibfield{author}{\bibinfo{person}{Jiafeng Guo}, \bibinfo{person}{Yixing Fan},
  \bibinfo{person}{Qingyao Ai}, {and} \bibinfo{person}{W~Bruce Croft}.}
  \bibinfo{year}{2016}\natexlab{}.
\newblock \showarticletitle{A deep relevance matching model for ad-hoc
  retrieval}. In \bibinfo{booktitle}{{\em Proceedings of the 25th ACM
  International on Conference on Information and Knowledge Management}}. ACM,
  \bibinfo{pages}{55--64}.
\newblock


\bibitem[\protect\citeauthoryear{Le and Mikolov}{Le and Mikolov}{2014}]%
        {le2014distributed}
\bibfield{author}{\bibinfo{person}{Quoc Le} {and} \bibinfo{person}{Tomas
  Mikolov}.} \bibinfo{year}{2014}\natexlab{}.
\newblock \showarticletitle{Distributed representations of sentences and
  documents}. In \bibinfo{booktitle}{{\em Proceedings of the 31st International
  Conference on Machine Learning (ICML-14)}}. \bibinfo{pages}{1188--1196}.
\newblock


\bibitem[\protect\citeauthoryear{Lin, Gao, Hu, Tang, Xiao, and Liu}{Lin
  et~al\mbox{.}}{2014}]%
        {lin2014mathematics}
\bibfield{author}{\bibinfo{person}{Xiaoyan Lin}, \bibinfo{person}{Liangcai
  Gao}, \bibinfo{person}{Xuan Hu}, \bibinfo{person}{Zhi Tang},
  \bibinfo{person}{Yingnan Xiao}, {and} \bibinfo{person}{Xiaozhong Liu}.}
  \bibinfo{year}{2014}\natexlab{}.
\newblock \showarticletitle{A mathematics retrieval system for formulae in
  layout presentations}. In \bibinfo{booktitle}{{\em Proceedings of the 37th
  international ACM SIGIR conference on Research \& development in information
  retrieval}}. ACM, \bibinfo{pages}{697--706}.
\newblock


\bibitem[\protect\citeauthoryear{Mikolov, Chen, Corrado, and Dean}{Mikolov
  et~al\mbox{.}}{2013a}]%
        {mikolov2013efficient}
\bibfield{author}{\bibinfo{person}{Tomas Mikolov}, \bibinfo{person}{Kai Chen},
  \bibinfo{person}{Greg Corrado}, {and} \bibinfo{person}{Jeffrey Dean}.}
  \bibinfo{year}{2013}\natexlab{a}.
\newblock \showarticletitle{Efficient estimation of word representations in
  vector space}.
\newblock \bibinfo{journal}{{\em arXiv preprint arXiv:1301.3781\/}}
  (\bibinfo{year}{2013}).
\newblock


\bibitem[\protect\citeauthoryear{Mikolov, Sutskever, Chen, Corrado, and
  Dean}{Mikolov et~al\mbox{.}}{2013b}]%
        {mikolov2013distributed}
\bibfield{author}{\bibinfo{person}{Tomas Mikolov}, \bibinfo{person}{Ilya
  Sutskever}, \bibinfo{person}{Kai Chen}, \bibinfo{person}{Greg~S Corrado},
  {and} \bibinfo{person}{Jeff Dean}.} \bibinfo{year}{2013}\natexlab{b}.
\newblock \showarticletitle{Distributed representations of words and phrases
  and their compositionality}. In \bibinfo{booktitle}{{\em Advances in neural
  information processing systems}}. \bibinfo{pages}{3111--3119}.
\newblock


\bibitem[\protect\citeauthoryear{Zanibbi, Aizawa, Kohlhase, Ounis, Topi{\'c},
  and Davila}{Zanibbi et~al\mbox{.}}{2016}]%
        {zanibbi2016ntcir}
\bibfield{author}{\bibinfo{person}{Richard Zanibbi}, \bibinfo{person}{Akiko
  Aizawa}, \bibinfo{person}{Michael Kohlhase}, \bibinfo{person}{Iadh Ounis},
  \bibinfo{person}{G Topi{\'c}}, {and} \bibinfo{person}{K Davila}.}
  \bibinfo{year}{2016}\natexlab{}.
\newblock \showarticletitle{NTCIR-12 MathIR task overview}.
\newblock \bibinfo{journal}{{\em NTCIR, National Institute of Informatics
  (NII)\/}} (\bibinfo{year}{2016}).
\newblock


\bibitem[\protect\citeauthoryear{Zhai and Lafferty}{Zhai and Lafferty}{2001}]%
        {zhai2001study}
\bibfield{author}{\bibinfo{person}{Chengxiang Zhai} {and} \bibinfo{person}{John
  Lafferty}.} \bibinfo{year}{2001}\natexlab{}.
\newblock \showarticletitle{A study of smoothing methods for language models
  applied to ad hoc information retrieval}. In \bibinfo{booktitle}{{\em
  Proceedings of the 24th annual international ACM SIGIR conference on Research
  and development in information retrieval}}. ACM, \bibinfo{pages}{334--342}.
\newblock


\end{thebibliography}
